\documentclass[aps,prc,twocolumn,floatfix,showpacs,a4paper,
nofootinbib,amsmath,amssymb]{revtex4-1}
\usepackage{graphicx}
\usepackage{dcolumn}
\usepackage{bm}
\usepackage{color}

\newcommand{\be}{\begin{equation}}
\newcommand{\ee}{\end{equation}}
\newcommand{\ba}{\begin{eqnarray}}
\newcommand{\ea}{\end{eqnarray}}
\newcommand{\bd}{\begin{displaymath}}
\newcommand{\ed}{\end{displaymath}}
\renewcommand{\vec}[1]{\mbox{\boldmath$#1$}}

\begin{document}
\title{New method to detect rotation in high energy heavy ion collisions}

\author{ L.P. Csernai$^1$, S. Velle$^1$, and D.J. Wang$^{1,2}$  }

\affiliation{
$^1$ Institute of Physics and Technology, University of Bergen,
Allegaten 55, 5007 Bergen, Norway\\
$^2$ Key Laboratory of Quark and Lepton Physics (MOE) and Institute of Particle Physics,
Central China Normal University, Wuhan 430079, China.
}

\begin{abstract}
With increasing beam energies the angular momentum of the fireball
in peripheral heavy ion collisions is increasing, and the proposed
Differential Hanbury Brown and Twiss analysis is able to estimate
this angular momentum quantitatively. The method detects specific
space-time correlation patterns, which are connected to rotation.
\end{abstract}

\date{\today}


\maketitle

\section{Introduction}

In high energy peripheral heavy ion collisions the high angular momentum 
is realized in rotating flow,
large velocity shear, vorticity and circulation. Viscous, explosive expansion
leads to the decrease of vorticity and circulation with time, however, with
small viscosity the vorticity remains significant at the final freeze-out
(FO) stages. The proposed Differential Hanbury Brown and Twiss (HBT) analysis,
a combination of standard two particle correlation functions,
is adequate to analyze rotating systems. At the present collision energies
the angular momentum and rotation is becoming a dominant feature of reaction
dynamics, and up to now the rotation of the system was never analysed,
neither with the HBT method nor in any other way.
We present and analyse this method and its results
in a high resolution, Particle In Cell fluid dynamics model.
Fluid dynamics is proven to be the best theoretical method to describe 
collective flow phenomena.
The same model was used to predict the
rotation in peripheral ultra-relativistic reactions
\cite{hydro1},
to point out the possibility of Kelvin Helmholtz Instability (KHI)
\cite{hydro2},
flow vorticity
\cite{CMW12} 
and polarization arising from local rotation, i.e. vorticity
\cite{BCW2013}.
The model was also tested for its numerical viscosity and the
resulting entropy production
\cite{Horvat}.
The formation of KHI was also observed recently in AdS/CFT holography,
where the instability is even more pronounced in peripheral reactions, 
although the time scale is sufficiently short only at high
quark chemical potentials as at FAIR, NICA and RHIC-BES
\cite{Mcinnes14}.

The total angular momentum of the fireball is maximal at
$b=0.3 b_{max}$ \cite{VAC13}, while the angular momentum per
net baryon charge is maximal around $b=(0.5 -0.8) b_{max}$. 
At ultra-peripheral
collisions fluctuations dominate collective effects.
According to the present analysis the Differential HBT method is indicating
rotation via particles at collective momenta, $p_t \approx (0.5 - 2)$ GeV/c
the best, and the magnitude of the introduced 
Differential Correlation Function
is monotonically increasing with the angular momentum.

%%%%%%%%%%%%%%%%%%%%%%%%%%%%%%%%%%%
\section{Correlation Function from Fluid Dynamics}
%%%%%%%%%%%%%%%%%%%%%%%%%%%%%%%%%%%

The pion correlation function is defined as the inclusive two-particle
distribution divided by the product of the inclusive one-particle
distributions, such that \cite{WF10}:

\begin{equation}
C( p_1 , p_2) =
\frac{P_2( p_1, p_2)}{P_1( p_1)P_1( p_2)},
\end{equation}

where $p_1$ and $p_2$ are the 4-momenta of pions.
We introduce the center-of-mass momentum
\footnote{The vector $\vec k$ is the wavenumber vector,
$ \vec k = \vec p/\hbar$ so for numerical calculations we have to
use that $\hbar c =$ 197.327 MeV fm., The same applies to $\vec q$.}
:
$
 k = \frac{1}{2} ( p_1 +  p_2) \ ,
$
and the relative momentum
$
 q =  p_1 -  p_2 \ ,
$
where from the mass-shell condition \cite{WF10} $q^0 = \vec k \vec q / k^0$.
We use a method for moving sources presented in Ref. \cite{Sinyukov-1}.
\begin{equation}
C(k,q) =
1 + \frac{R(k,q)}{\left| \int d^4 x\,  S(x, k) \right|^2}\ ,
\label{C-def}
\end{equation}

where

\be
\begin{split}
R(k,q) =&  \int d^4 x_1\, d^4 x_2\, \cos[q(x_1-x_2)] \times\\
& S(x_1,  {k}+{q}/2)\, S(x_2,{k}-{q}/2)\ .
\end{split}
\label{R1}
\ee
Using the emission function $S(x,k)$, discussed in refs.
\cite{CS13},
here $R(k,q)$ can be calculated
\cite{Sinyukov-1}
via the function

\be
J(k,q) = \int d^4x\ S(x,k+q/2)\, \exp(iqx) \ ,
\label{J-def}
\ee

and we obtain the $R(k,q)$ function as
$ R(k,q) = Re\, [ J(k,q)\, J(k,-q) ] $.

We estimate the local pion density by the specific entropy, $\sigma(x)$, as
\ $n_\pi(x) \propto n(x)\,\sigma(x)$, \ where \ $n(x)$ \ is the proper
net baryon charge density. 
\footnote{
At the latest times presented here, $t=3.56$ fm/c, ($\sim 8$ fm/c after
the initial touch) the net baryon density is sufficiently large at 
non-vanishing entropy, so this approximation is satisfactory. At later times
the entropy density becomes dominant, while the net baryon density decreases".
}
Thus the local invariant pion density is given
by the J\"uttner distribution as

\be
f^J(x,p) = \frac{n(x) \sigma(x)}{C_\pi}
\exp\left(-\frac{p^\mu u_\mu (x)}{T(x)}\right),
\label{Jut-2}
\ee
where \ $C_\pi=4\pi m_\pi^2T K_2(m_\pi/T)$, \ at temperature $T$,
and $K_2$ is a modified Bessel function.

We assume that the single particle distributions, $f(x,p)$, in
the source functions are J\"uttner distributions, which depend on the
local velocity, $u^\mu(x)$, and
we use the notation $u_1 = u(x_1) = u^\mu(x_1)$.

By using the Cooper-Frye (CF) freeze out description we can connect
the Source function, $S$, to the phase space distribution function
on the freeze out hypersurface.
Let the space-time points of the hyper-surface be given in parametric 
form $x_{FO} = x_{FO}(x)$,
which can be given by the freeze out condition 
(e.g $t=$const., $\tau=$const., $T=$const. or other).
In the source function formalism this corresponds to a 4-volume
integral 
\be
\begin{split}
& \int d^4x\ S(x,p) = \int d^4x\  f^J(x,p)\ P(x,p) =  \\
& \int d^4x\  f^J(x,p)\ \delta(x-x_{FO})\ p^\mu \hat{\sigma}_\mu \ ,
\end{split}
\nonumber
\ee
where the emission probability is \cite{Cso-5} 
$P(x,p) = \delta(x-x_{FO})\ p^\mu \hat{\sigma}_\mu$.
This CF freeze out treatment is the most frequent in fluid 
dynamical models. This sudden freeze out assumption can be 
relaxed by assuming an extended freeze out layer in the
space time via replacing the Dirac delta function with a
freeze out profile function in $P(x,p)$, e.g.:
\be
\begin{split}
& P(x,p) = 
\delta(x-x_{FO})\ p^\mu \hat{\sigma}_\mu   \longrightarrow \\
& \frac{1}{\sqrt{\Delta \pi}} 
\exp\left(-\frac{(s-s_{FO})^2}{\Delta}\right)\
p^\mu \hat{\sigma}_\mu \ ,
\end{split}
\nonumber
\ee  
where $s$ is a local coordinate in the direction of $\hat{\sigma}^\mu$,
and the local width of the freeze out layer is $\Delta =
\Delta(x)$ (which should tend to zero to get the Dirac delta function
for the emission probability). This description is then completely general,
with the only assumption that the emission probability has a 
Gaussian profile. (This could also be relaxed.)

If we assume that the two coincident particles originate from two
points, $x_1$ and $x_2$,
the expression of the correlation function, Eq. (\ref{R1}) will
be become \cite{CS13}
\be
\begin{split}
R(k,q) = \int\!  &  d^4 x_1 d^4 x_2\, S(x_1, k) S(x_2, k)
    \cos[q(x_1{-}x_2)] \times  \\
&   \exp\left[ -\frac{q}{2} \cdot \left( \frac{u(x_1)}{T(x_1)}
                                  -\frac{u(x_2)}{T(x_2)}
              \right)\right],
\end{split}
\label{R-def2}
\ee
and the corresponding $J(k,q)$-function is
\be
J(k,q) =  \int d^4x\ S(x,k)\,
\exp\left[ - \frac{q \cdot u(x)}{2 T(x)} \right]\, \exp(iqx)\ ,
\label{J2}
\ee

In Ref. \cite{CS13} different one, two and four source systems
were tested with and without rotation.
Here we study only the case where the emission is
{\it asymmetric} and dominated by the fluid elements facing the detector.

In numerical fluid dynamical studies of symmetric (A+A) nuclear
collision the initial state is symmetric around the center of mass
(c.m.) of the system, and (if we do not consider random fluctuations)
this symmetry is preserved during the fluid dynamical evolution.

Let us consider the usual conventions, $z$ is the beam axis, and the
positive $z$-direction is the direction of the projectile beam. The
impact parameter vector points into the positive $x$-direction, i.e.
towards the projectile. Finally the $y$-axis is orthogonal to both.

The fluid dynamical system, without fluctuations
can be considered as a set of symmetric pairs of fluid cells.

The emission probabilities
from the two fluid cells of a source pair are not equal.

If we have several fluid cell sources, $s$, with Gaussian space-time (ST)
emission profiles, then the source function
in J\"uttner approximation is
\be
\begin{split}
& \int d^4x\, S({x}, k) = \sum_s \int_s d^3x_s\, dt_s\, S(x_s,k) = \\
& (2\pi R^2)^{3/2}\,
\sum_s  \frac{\gamma_s n_s(x) \, (k_\mu\, \hat\sigma^\mu_s)}{C_s} \,
\exp\left[-\frac{k \cdot u_s}{T_s}\right] \ ,
\end{split}
\ee
\vskip -4mm
\noindent
where $n_s=n_{\pi}$, and the spatial integral over a cell volume is,
$V_{cell}=(2\pi R^2)^{3/2}$ while the time integral is
normalized to unity. Similarly the $J$-function is
\be
J(k,q) = \sum_s
   e^{- \frac{q}{2} \cdot\frac{u_s}{T_s}}\, e^{iqx_s}
   \int_s d^4x\ S_s(x,k)\, e^{iqx}\ .
\ee
We then assume that the FO layer is relatively narrow compared
to the spatial spread of the fluid cells, so that the peak emission times,
$t_s$, of all fluid cells are the same. Then the $\exp(iq^0 t_s)$
factor drops out from the $J(k,q) J(k,-q)$ product.
\footnote{If the emission is happening through
a layer with time-like normal, but the peak is not at constant $t_s$, but
rather at constant $\tau_s$, then we can adapt the coordinate
system accordingly, i.e. we can use the $\tau, \eta$ coordinates instead
of $t, z$, see e.g. \cite{Cso-5}.}
This FO simplification is justified for
rapid and simultaneous hadronization and FO from
the plasma. For dilute and transparent matter the correlations
from the time dependence of FO should be handled the same
way as the spatial dependence.

Due to mirror symmetry with respect to the $[x,z]$,
reaction plane, it is sufficient to describe the
cells on the positive side of the $y$-axis. The other
side is the mirror image of the positive side. Then we
can evaluate the correlation function the same way as in Ref.
\cite{CS13}.

Thus we define the quantities:
\be
\begin{split}
& Q_c = \left(2 \pi R^2 \right)^{3/2}
    \exp\left[-\frac{R^2  q^2}{2}\right], \\
& P_s = \frac{\gamma_s n_s}{C_s}
    \exp\left[-\frac{ k_0\,u^0_s}{T_s}\right],  \\
& Q_s^{(q)} = \exp\left[-\frac{q_0\,u^0_s}{2T_s}\right],  \\
& w_s = (k_\mu\, \hat\sigma^\mu_s) \ \exp\left[-\frac{\Theta_s^2}{2}
        q_0^2\right] \ , \\
\end{split}
\label{Qw-s}
\ee
where $u^0_s = \gamma_s$,
the local 4-direction normal of the mean particle  emission from an
ST point of the flow is $\hat\sigma^\mu_s$ (assumed to be time-like),
$R$ is the size (radius) of the fluid cells, and
$\Theta_s$ is the path length of the time integral from the
ST point of the source, $s$,
while assuming a Gaussian emission time profile \cite{CS13}.
The weights, $ w_s$ arise directly from the Cooper-Frye formula \cite{Cso-5}.

We can reassign the summation for pairs, so that $s = \{i,j,k\}$ will
correspond to a pair of cells: at $\{i,j,k\}$ and its reflected
pair across the c.m. point at the same time at $\{i^*,j^*,k^*\}$.
Then the function $S(k,q)$ becomes
\be
\begin{split}
& \int d^4x S(x,k) = \left(2 \pi R^2 \right)^{3/2} \times\\
& \sum_s P_s \left[
   w_s   \exp\left( \frac{\vec k \vec u_s  }{T_s}\right)
 + w_s^* \exp\left( \frac{\vec k \vec u_s^*}{T_s}\right) \right] ,
\end{split}
\label{Spair}
\ee
while, the function $J(k,q)$ becomes
\be
\begin{split}
& J(k,q) = Q_c \sum_s P_s
 \left[
  Q_s^{(q)} w_s  \exp\left[\left(\vec k{+}\frac{\vec q}{2}\right)\!
      \frac{\vec u_s}{T_s}\right]
           e^{i\vec q \vec x_s}  \right. \\
& \left. +
  Q_s^{(q)} w_s^* \exp\left[\left(\vec k{+}\frac{\vec q}{2}\right)\!
       \frac{\vec u_s^*}{T_s}\right]
           e^{i\vec q\vec x_s^*} \right]
\end{split}
\nonumber
\label{Jpairs}
\ee
\vskip -4mm
Only the mirror symmetry across the participant c.m. is assumed, which is
always true for globally symmetric, A+A, heavy ion collisions in a
non-fluctuating fluid dynamical model calculation.
Then the correlation function can be evaluated using
Eqs. (\ref{C-def}-\ref{J-def}).

By using few fluid cell sources for tests, in Ref. \cite{CS13} it was shown 
that in case of a globally symmetric fluid dynamical configuration
the correlation function
only includes $\cos(c\, \vec k \vec u_s)$ and 
$\cosh(c\, \vec k \vec u_s)$ terms,
therefore it will not depend on
the {\em direction} of the velocity, only on its magnitude.
The {\em direction} dependence becomes apparent in the correlation
function only if we take into account that due to
the radial expansion and the opacity of the strongly interacting QGP,
the emission probability from the far side of the system
is reduced compared to the side of the system facing the detector.

Based on the few source model results the Differential HBT method
was  introduced by evaluating the difference of two correlation
functions measured at two symmetric angles, forward and backward
shifted in the reaction plane in the participant c.m. frame
by the same angle, i.e. at $\eta = \pm $const., so that the
Differential Correlation Function (DCF) becomes
\be
\Delta C(k,q) \equiv C(k_+,q_{out}) - C(k_-,q_{out}).
\ee
For the exactly $\pm x$ -symmetric spatial configurations
(i.e. $k_{+x} = k_{-x}$ and  $k_{+z} = - k_{-z}$), e.g. central collisions
or spherical expansion, $\Delta C(k,q)$ would vanish!
It would become finite if the rotation introduces an asymmetry.

%%%%%%%%%%%%%%%%%%%%%%%%%%%%%%
\section{The Freeze-out}
%%%%%%%%%%%%%%%%%%%%%%%%%%%%%%
%
The HBT method is sensitive to the time development of particle 
emission, and well suited to transport models where emission
happens during the ST history of the collision, although 
the emission is concentrated at a FO layer. 
\begin{figure}[ht] %%%%%%%%%%
\begin{center}
      \includegraphics[width=7.6cm]{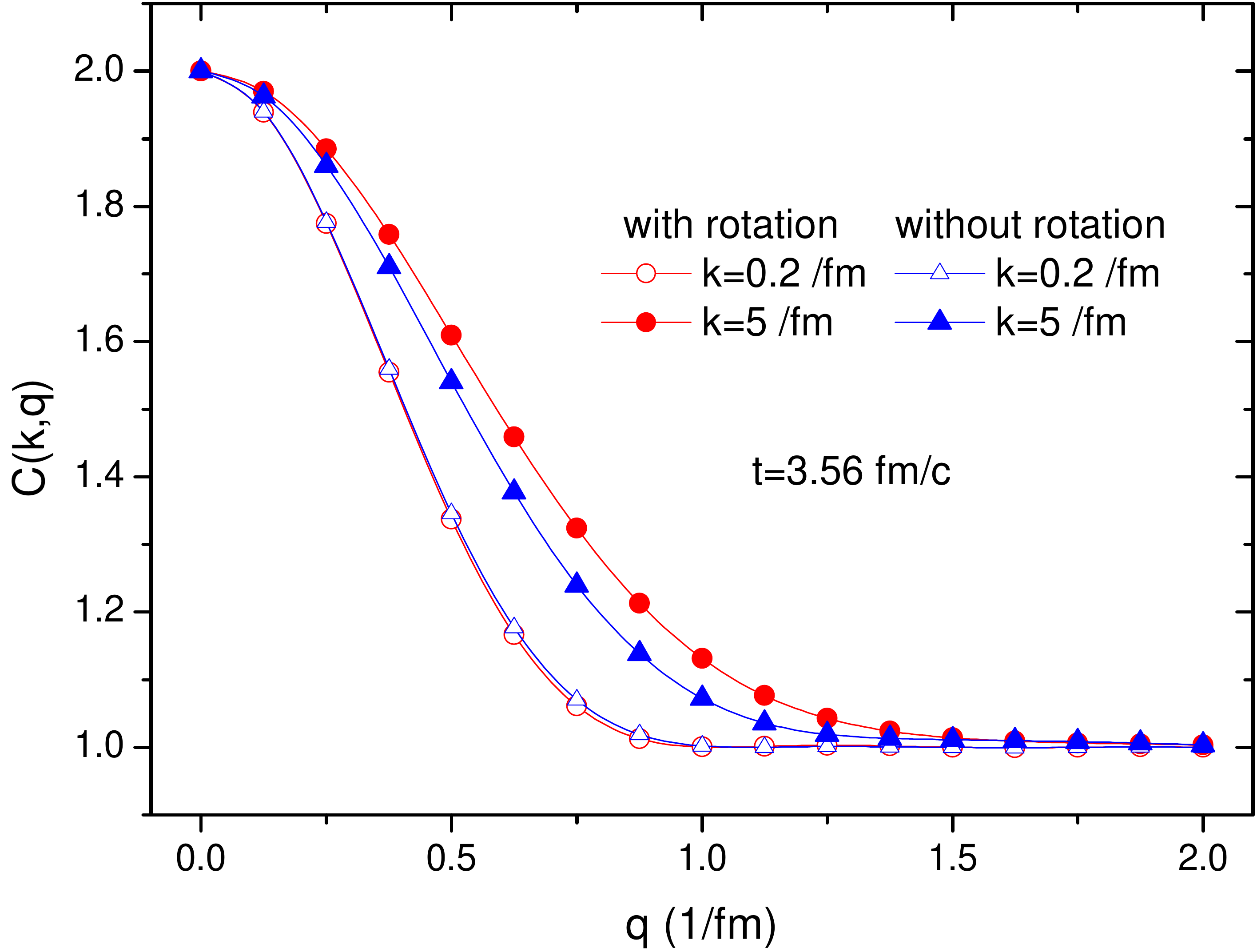}
\end{center}
\vskip -0.6cm
\caption{(color online)
The dependence of the standard correlation function in the $\vec k_+$
direction from the collective flow, at the final time, 3.56 fm/c 
after reaching local equilibrium and 8.06 fm/c from the first touch,
including the initial longitudinal expansion Yang-Mills field dynamics
\cite{magas}.
%C-wwo-flow-t=3.56
}
\label{Fig-1}
\end{figure}
The fluid dynamical model
is also able to describe emission in a ST volume or layer 
\cite{M-2,M-3}, or in hybrid models where a microscopic transport module 
is attached to the fluid dynamics, e.g. \cite{Yan12}. 
The determination of the FO surface normal or the
emission direction from the ST FO layer and the emission
profile in this layer are the subjects of present theoretical
research, see \cite{AVC13,VAC13,Yun10,HP12,AVY13}.
This complex FO process certainly has an influence on the HBT analysis, 
but our present aim is not to reproduce exactly a given 
experiment.

We focus on a single collective effect, the rotation, developing
from the angular momentum during the initial stages of the
fluid dynamics. Thus we constrain our discussion 
to the fluid dynamical stage, and adopt a relatively simple 
FO description from ref. \cite{Cso-5}, which can be implemented in  
Eq. (\ref{Qw-s}). This provides the
emission weight factors, $w_s$, which depend on the local mean 
emission direction $\hat\sigma^\mu$, and
the flow velocity at the emission point.

The detector configuration is given by the two particles
reaching a given detector in the direction of $\vec k$. Thus
the emission weights depend on the normal of the emission surface 
and of the emission, i.e. $\hat\sigma$ and $\hat k$.
Most of the particles FO in a layer along a
constant proper time hyperbola, with a dominant flow 4-velocity
normal to this hyperbola: $\hat\sigma^\mu \approx u^\mu$.
The origin of the hyperbola is at a ST point, which may precede 
the impact of the Lorentz contracted nuclei \cite{AVC13}.

We assume in the actual numerical calculations that in the expression
of the weight, in Eq. (\ref{Qw-s}),
is the same for all surface layer elements:
$Q_s^{(q)} = Q^{(q)}$ and $\Theta_s = \Theta$, so that
$
 w_s = (k_\mu\,\hat\sigma^\mu_s)\
\exp( -\Theta^2 q_0^2 / 2 ) \ ,
$
where
$ \hat \sigma_{s\mu} = (\sigma_s^0, \vec \sigma_s) $,
so that
$ \  k_\mu\, \hat\sigma_s^\mu =
  k^0 \sigma_s^0 + \vec k \vec \sigma_s \ .
$
If the emission path time-length, $\Theta$, tends to zero,
then the time modifying factor becomes unity.
With the choice $\hat\sigma_\mu = u_\mu$, the time-like FO normal is
$\hat\sigma_{s\mu} = (\gamma_s, \vec u_s)$. Then
$(k_\mu \hat\sigma^\mu_s) = \gamma_s k_0 {+} \vec k \vec u_s $.
So the weight becomes
\be
 w_s = ( \gamma_s k_0 {+} \vec k \vec u_s) \
        \exp(-\Theta^2 q_0^2 / 2) .
\label{ws}
\ee
\noindent
This weight is explicitly different for the mirror image cell at
$\vec x_s^* \rightarrow -\vec x_s $, where
$\vec u_s^* \rightarrow -\vec u_s $ and then
$
 w_s^*~=~(\gamma_s k_0 {-} \vec k \vec u_s)
        \exp(-\Theta^2 q_0^2 / 2)\ .
$

The weight factors appear both in the nominator and denominator
of the correlator, so its normalization is balanced. On the
other hand the role of the different factors in the weight
have an effect to determine, which cells contribute more, 
which cells contribute less to the integrated
result.

\begin{figure}[h] %%%%%%%%%%
\begin{center}
      \includegraphics[width=7.6cm]{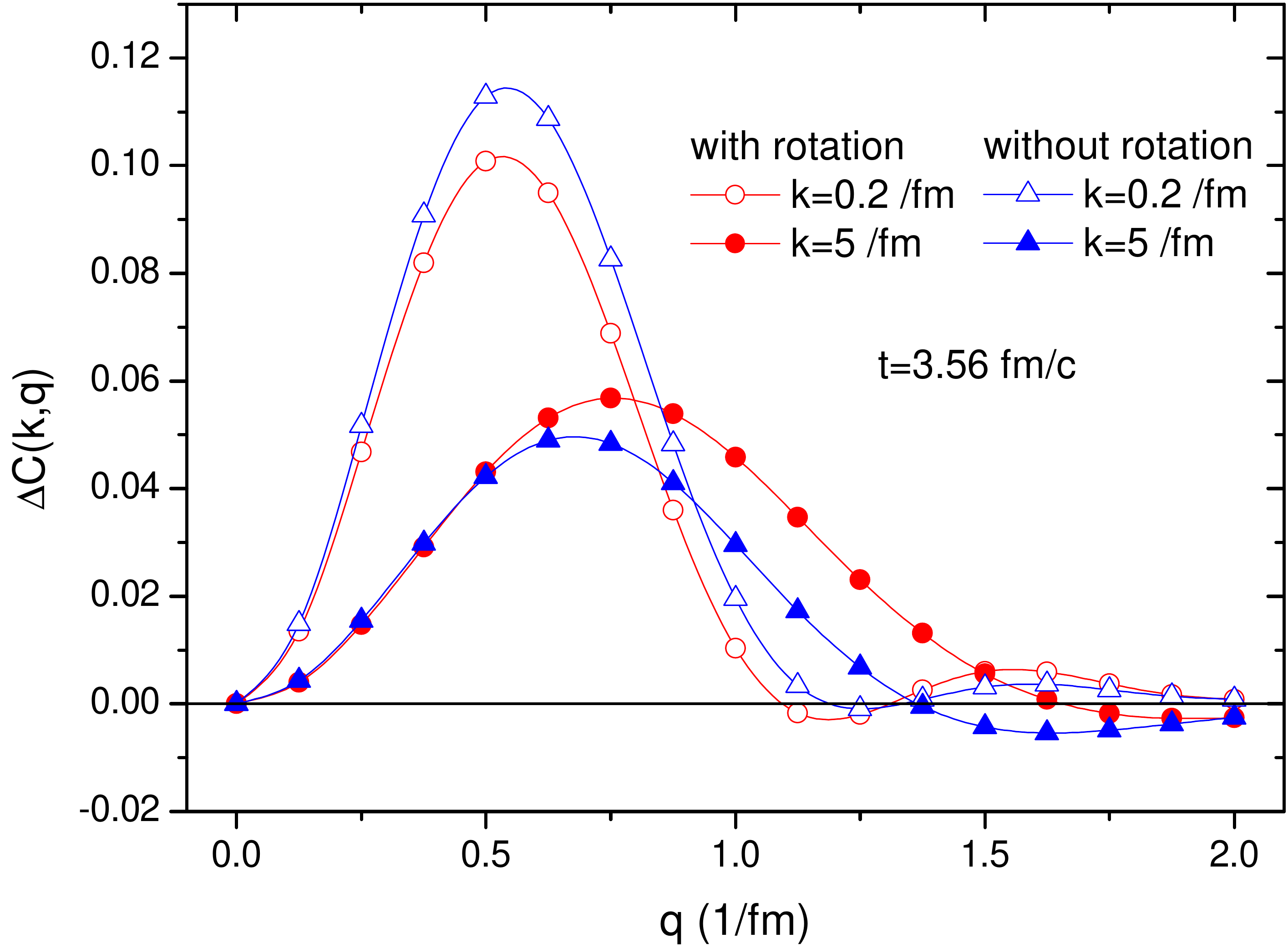}
\end{center}
\vskip -0.3cm
\caption{(color online)
The Differential Correlation Function, $\Delta C(k,q)$,
at the final time with and without rotation.
}
\label{Fig-2}
\end{figure}
%

%%%%%%%%%%%%%%%%%%%%%%%%%%%%%%%%%%%
\section{Results}
%%%%%%%%%%%%%%%%%%%%%%%%%%%%%%%%%%%
%
The sensitivity of the standard correlation function on the 
fluid cell velocities decreases with decreasing distances
among the cells. So, with a large number of densely placed fluid cells
where all fluid cells contribute equally to the correlation function,
the sensitivity on the flow velocity becomes negligibly weak.

Thus, the emission probability from different ST regions of the
system is essential in the evaluation. This emission asymmetry
due to the local flow velocity
occurs also when the FO surface or layer is isochronous
or if it happens at constant proper time.

We studied the fluid dynamical patterns of the calculations
published in Ref. \cite{hydro2},
where the appearance of the KHI is discussed under different conditions.
We chose the configuration, where both the rotation \cite{hydro1},
and {\bf the KHI occurred}, at
$b=0.7 b_{max}$ with high cell resolution and low numerical
viscosity at LHC energies, where the angular momentum is large,
$L \approx 10^6 \hbar$ \cite{VAC13}.

In spatially symmetric few source configurations \cite{CS13},
the standard correlation function
did not show any difference if it is measured at two symmetric
$\vec k$ and $\vec q$-out angles, e.g. in the reaction, [x-z] plane at
$\vec k_+ = (k_x, 0, +k_z)$,  $\vec q_+ = (q_x, 0, +q_z)$ and
$\vec k_- = (k_x, 0, -k_z)$,  $\vec q_- = (q_x, 0, +q_-)$,
i.e. $\Delta C(k,q)$ vanished.
Here we have chosen two directions at $\eta = \pm 0.76$, that is
at polar angles of $90\pm40$ degrees. These are measurable with
the ALICE TPC and at the ATLAS and CMS also.

The standard correlation function is both influenced by the
ST shape of the emitting source as well as its velocity
distribution. The correlation function becomes narrower in $q$
with increasing time primarily due to the rapid expansion of the
system. At the initial configuration the increase of $|\vec k|$
leads to a small increase of the width of the correlation function.

Nevertheless, in theoretical models we can switch off the rotation component
of the flow,
and analyse how the rotation influences the correlation function
and especially the DCF, $\Delta C(k,q)$.

Fig. \ref{Fig-1} compares the standard correlation functions
with and without the rotation component of the  flow at the final time moment.
Here we see that the rotation leads to a small increase of the
width in $q$ for the distribution at high  values of $|\vec k|$, while at 
low momentum there is no visible difference.

In Fig.\,\ref{Fig-2}
$\Delta C(k,q) $ is shown for the configuration with and without
rotation. For $k=5/$fm the rotation increases both the 
amplitude  and the width of $\Delta C$. 
The dependence on $|\vec k|$ is especially large at the final time.

\begin{figure}[ht] %%%%%%%%%%
\begin{center}
      \includegraphics[width=4.00cm]{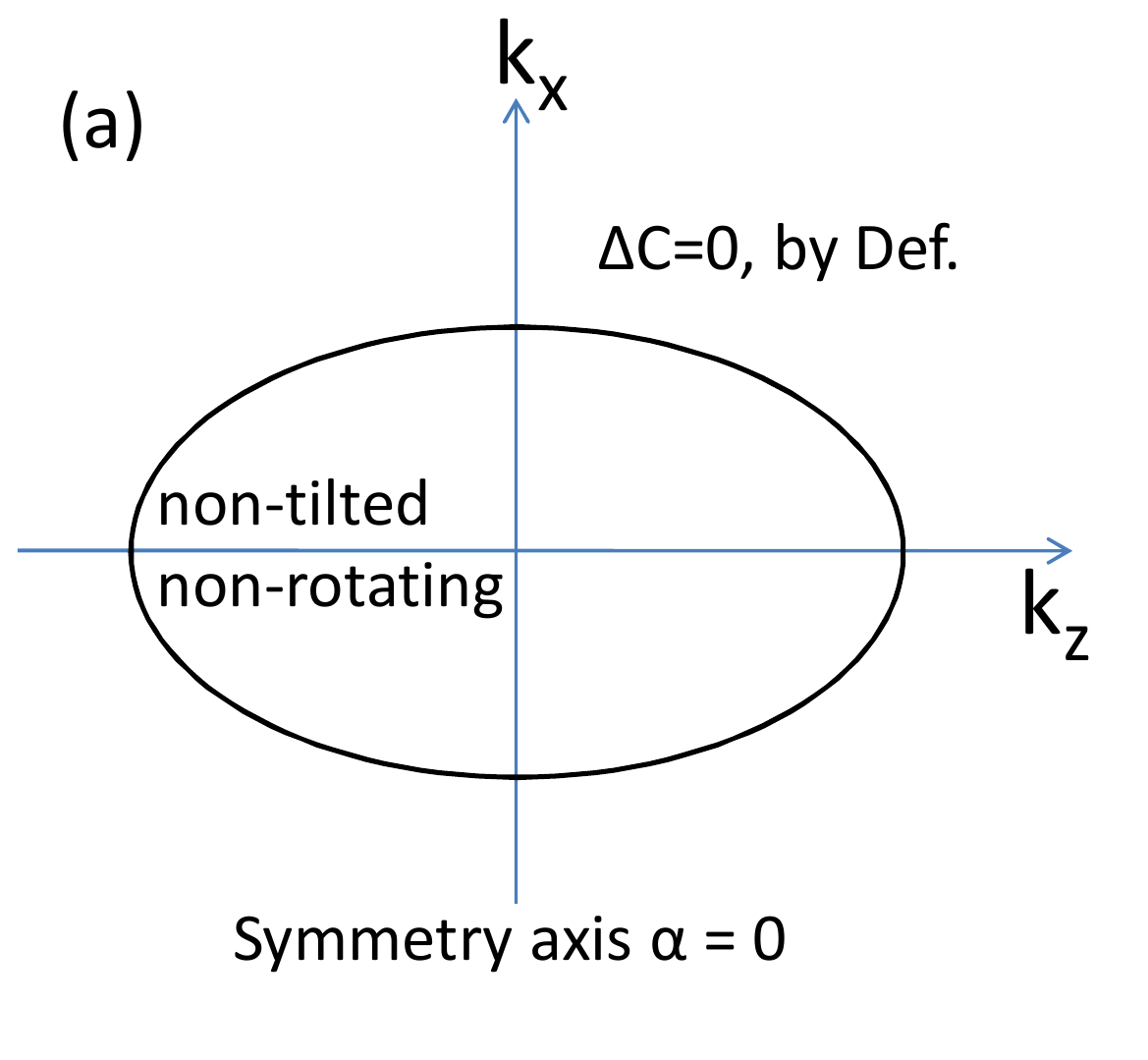}
      \includegraphics[width=4.00cm]{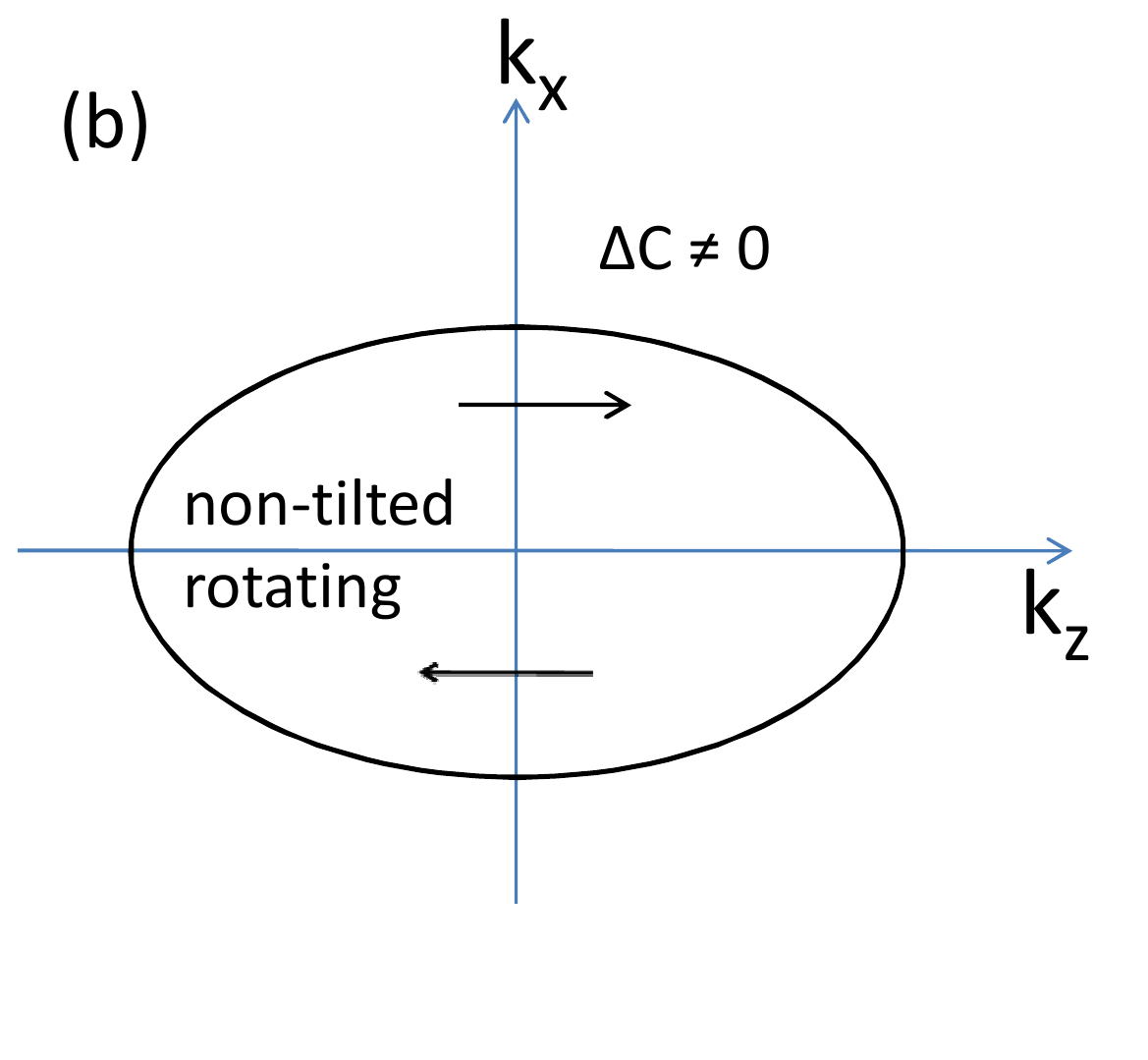}\\
      \includegraphics[width=4.00cm]{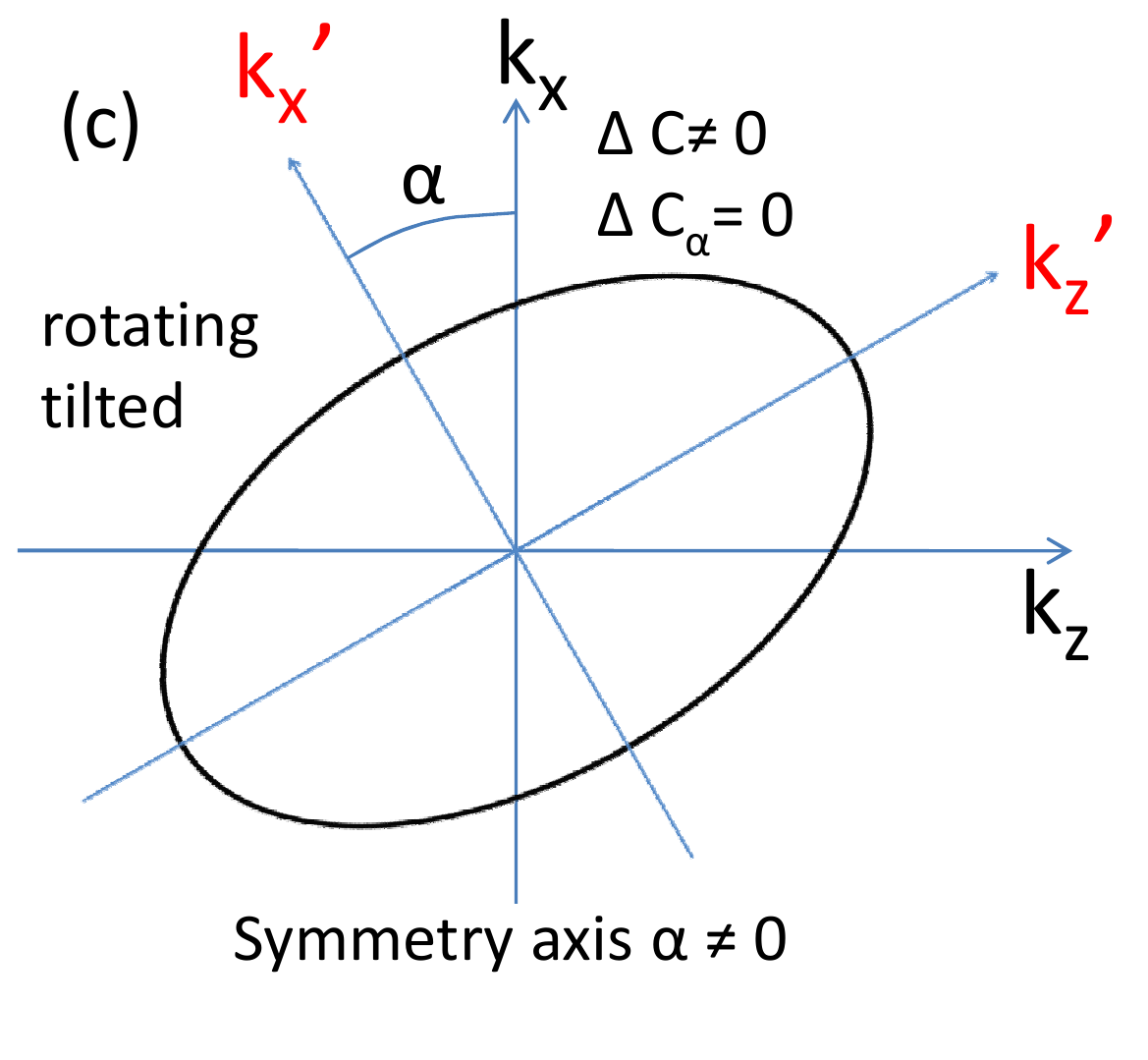}
      \includegraphics[width=4.00cm]{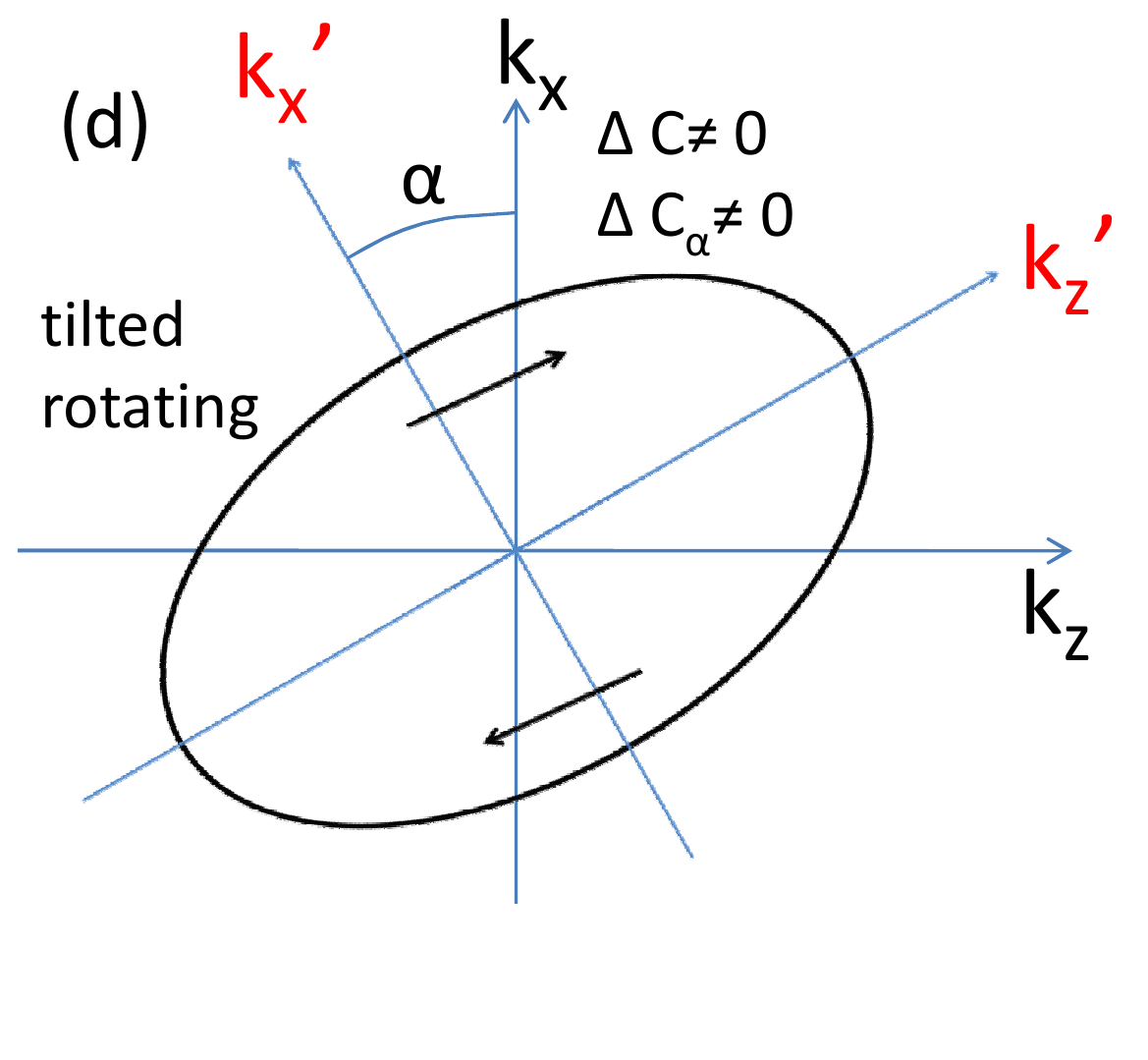}
\end{center}
\caption{
(Color online) 
Sketch of the configuration in different reference frames, with and
without rotation of the flow. The non-rotating configurations have
may have radial flow velocity components only. The DCF, 
$\Delta C_\alpha(k,q) $
is evaluated in a $K'$ reference frame rotated by and angle $\alpha$
in the $x,z$, reaction plane. We search for the angle $\alpha$,
where the non-rotataing configuration is "symmetric", so that it
has a "minimal" DCF as shown in Fig. \ref{Fig-3}.  
}
\label{FigS-3}
\end{figure}

In the original $K$ frame defined by the beam direction and the
impact parameter, we can describe the vector $\vec k$ with coordinates,
$ \vec k = \left\{ \begin{array}{c} k_x\\ k_z \end{array} \right\} $. 
In the $K'$ frame the
same vector is then
\be
\vec k'(\alpha) = 
\left\{ \begin{array}{c} k_{x'} \\ k_{z'} \end{array} \right\}
=
\left\{ \begin{array}{c} k_x \cos \alpha - k_z \sin \alpha \\
 k_z \cos \alpha + k_x \sin \alpha \end{array} \right\} \ .
\ee
Then one can define the DCF,
\be
\Delta C_\alpha(\vec k', \vec q' ) \ ,
\ee
which depends on the angle $\alpha$.
We have to find the proper symmetry axes of the emission ellipsoid. The 
conventional way would be the standard azimuthal HBT, however, we can 
exploit the features of the DCF. As the analytic examples \cite{CS13} 
show if (i) the shape is symmetric around the $x'$ axis, and (ii) there
is no rotation in the flow, then 
\be
\Delta C_\alpha\left(\vec k', \vec q'\right) = 0\ .
\ee
Thus we can use the DFC to find the angle, $\alpha'$, of the proper
frame $K'$ also. For a given $|\vec k|$ (e.g. $|\vec k|= 5/$fm), 
we search for the minimum of the norm of the DCF as a function of $\alpha$.

\begin{figure}[ht] %%%%%%%%%%
\begin{center}
      \includegraphics[width=7.6cm]{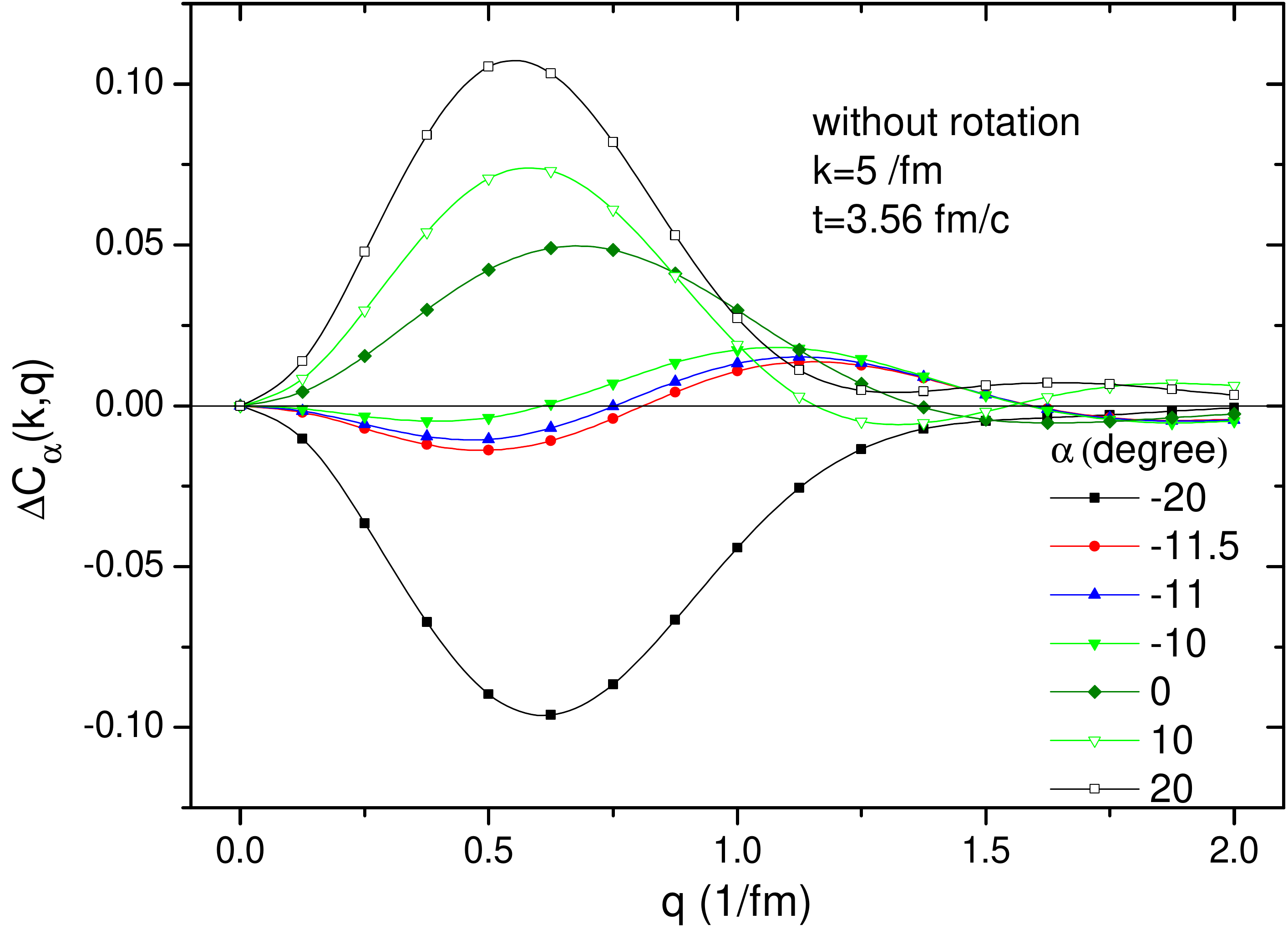}
\end{center}
\caption{
(Color online) 
The DCF at average pion wavenumber,
$k = 5/$fm and fluid dynamical evolution time, $t=3.56$fm/c, as a function
of the functions of momentum difference in the "out" direction $q$
(in units of 1/fm). The DCF is evaluated in a frame rotated in the
reaction plane, in the c.m. system by angle $\alpha$. 
}
\label{Fig-3}
\end{figure}

This procedure is performed and the result is shown in Fig. \ref{Fig-3}
We separated the effect of the rotation by finding the
symmetry angle where the rotation-less configuration yields vanishing 
or minimal $\Delta C$ for a given transverse momentum $k$.

The 
figure shows the result where the rotation component of the velocity
field is removed. The DCF shows a minimum in its integrated value
over $q$, for $\alpha = -11$ degrees.   The shape of the DCF changes 
characteristically with the angle $\alpha$. 
Unfortunately this is not possible experimentally, so the direction
of the symmetry axes should be found with other methods, like global flow 
analysis and/or azimuthal HBT analysis. 
To study the dependence on the angular momentum the same study was 
for lower angular momentum also, i.e. for a lower (RHIC) energy Au+Au
collisions at the same impact parameter and time. We identified the 
angle where the rotation-less DCF
was minimal, which was 
$\alpha = -8$ degrees, less than the deflection at higher angular momentum.

\begin{figure}[ht] %%%%%%%%%
\vskip 4mm
\begin{center}
      \includegraphics[width=7.4cm]{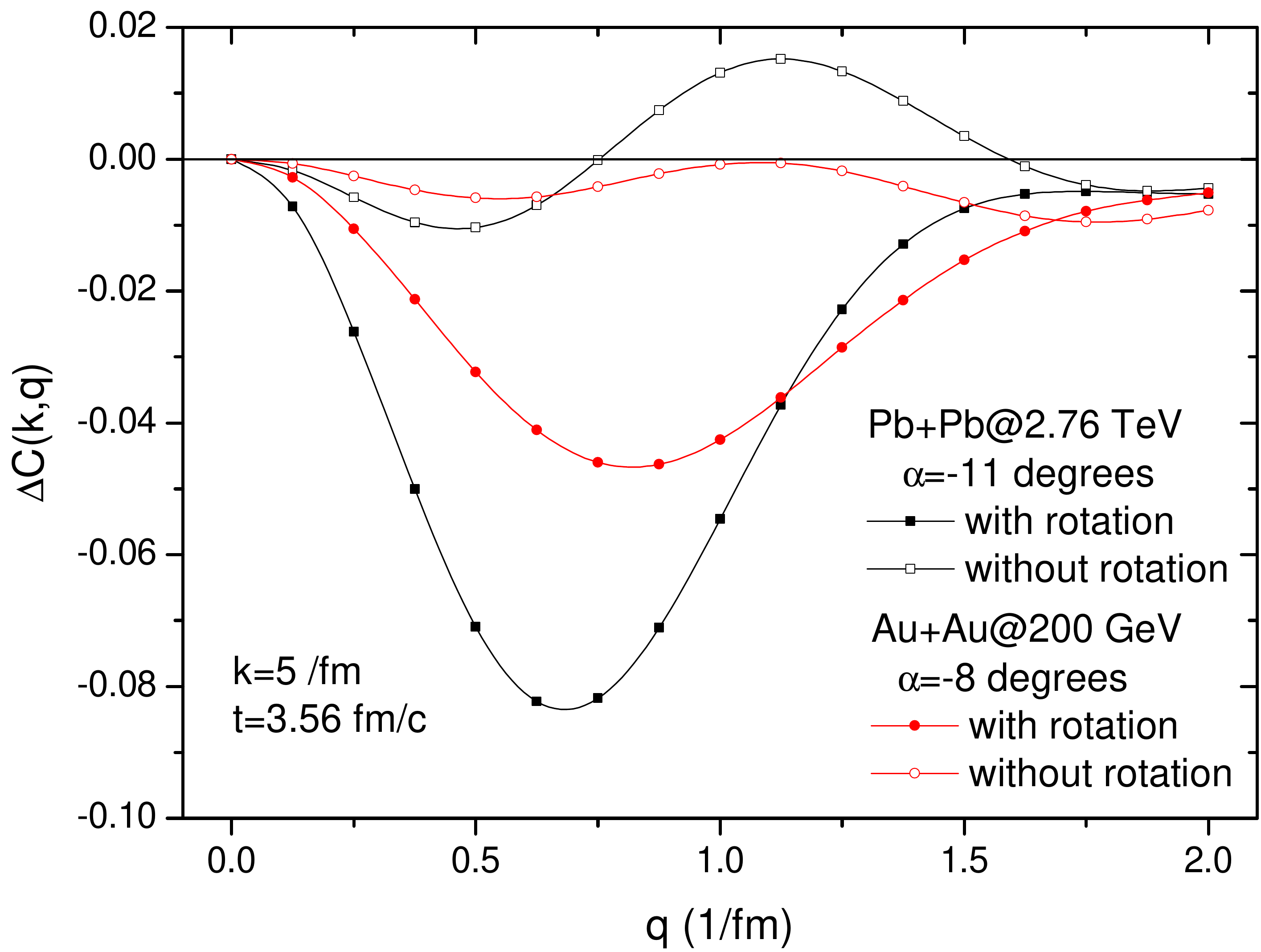}
\end{center}
\caption{(color online)
The DCF 
with and without rotation in the
reference frames, deflected by the angle $\alpha$, where the rotation-less
DCF is vanishing or minimal. In this frame the DCF of the original, rotating
configuration indicates the effect of the rotation only. 
The amplitude of the DCF 
of the original rotating configuration doubles for the higher energy (higher
angular momentum) collision.
}
\label{Fig-4}
\end{figure}

We did this for two different energies, Pb+Pb / Au+Au
at $\sqrt{s_{NN}}= 2.36 / 0.2$ TeV respectively, 
while all other parameters of the collision were the same. 
The deflection angle of the symmetry axis was $\alpha =-11/-8$ 
degrees\footnote{The negative angles are arising from the fact that
our model calculations predict rotation, with a peak rotated forward
\cite{hydro1}.}
respectively. In these deflected frames we evaluated $\Delta C$ for the
original, rotating configurations, which are shown in  Fig.\,\ref{Fig-4}.
This provides an excellent measure of the rotation.

On the other hand the rotation-less configuration cannot be generated from
experimental data in an easy way. Other methods like the
Global Flow Tensor analysis, or the azimuthal HBT analysis \cite{MLisa}
can provide an estimate
for finding the deflection angle  $\alpha$.

%%%%%%%%%%%%%%%%%%%%%%%%%%%%%%%%%%%
\section{Conclusion}
%%%%%%%%%%%%%%%%%%%%%%%%%%%%%%%%%%%
%
The analysed model calculations show that the Differential HBT analysis 
can give a good quantitative measure of the
rotation in the reaction plane of a heavy ion collision. These
studies show that this measure is proportional to the beam energy or total 
angular momentum (while the polarization \cite{BCW2013} is not).
It shows the best sensitivity at higher collective transverse momenta.

To perform the analysis in the rotation-less symmetry frame one can 
find the symmetry axis the best with the azimuthal HBT method, which
provides even the transverse momentum dependence of this axis \cite{MLisa}.

It is also important
to determine the precise Event by Event c.m. position of the
participants \cite{Eyyubova}, and minimize the effect of fluctuations
to be able to measure accurately the emission
angles, which are crucial in the present $\Delta C(k,q)$ studies.

\noindent
%%%%%%%%%%%%%%%%%%%%%%%%%%%%%%%%%%%
\section*{Acknowledgements}
%%%%%%%%%%%%%%%%%%%%%%%%%%%%%%%%%%%
This work was supported in part by the Helmholtz International Center for
FAIR. We thank F. Becattini, M. Bleicher, G. Graef, P. Huovinen and J. Manninen for
comments.

%%%%%%%%%%%%%%%%%%%%%%%%%%%%%%%%%%%%%%%


\begin{thebibliography}{99}
%%%%%%%%%%%%%%%%%%%%%%%%%%%%%%%%%%%%%%%



\bibitem{hydro1}
  L.P. Csernai, V.K. Magas, H. St\"ocker, and D.D. Strottman,
  Phys. Rev. C {\bf 84},  024914 (2011).

\bibitem{hydro2}
  L.P. Csernai, D.D. Strottman and C. Anderlik,
  Phys. Rev. C {\bf 85}, 054901 (2012).

\bibitem{CMW12}
%  Flow Vorticity in Peripheral High Energy Heavy Ion Collisions
   L.P. Csernai, V.K. Magas, and D.J. Wang,
   Phys. Rev. C 87, 034906 (2013). 
%  arXiv:1302.5310v1 [nucl-th]

\bibitem{BCW2013}
F. Becattini, L.P. Csernai, D.J. Wang,
Phys. Rev. C {\bf 88}, 034905 (2013).

\bibitem{Horvat}
  Sz. Horv\'at, V.K. Magas, D.D. Strottman, L.P. Csernai,
  Phys. Lett. B {\bf 692}, 277 (2010).

\bibitem{Mcinnes14}
  B. McInnes, and E.Teo, Nucl. Phys. B {\bf 878}, 186 (2014).

\bibitem{VAC13}
 %Longitudinal fluctuations of participant center-of-mass in heavy-ion coll..
 V. Vovchenko, D. Anchishkin, and L.P. Csernai, 
Phys. Rev. C {\bf 88}, 014901 (2013).

\bibitem{WF10}
  W. Florkowski: {\it Phenomenology of Ultra-relativistic
  heavy-Ion Collisions}, World Scientific Publishing Co., Singapore (2010).

\bibitem{Sinyukov-1}
 A.N. Makhlin, Yu.M. Sinyukov, Z. Phys. C {\bf 39}, 69 (1988);
 S.V. Akkelin, Yu.M. Sinyukov, Z. Phys. C {\bf 72}, 501 (1996);
 T. Cs\"org\H{o}, S.V. Akkelin, Y. Hama, B. Luk\'acs, and Yu.M. Sinyukov,
 Phys. Rev. C {\bf 67}, 034904 (2003) .

\bibitem{CS13}
L.P. Csernai, S. Velle, (2013) arXiv:1305.0385

\bibitem{Cso-5}
%Particle Interferometry from 40 MeV to 40 TeV
T. Cs\"org\H{o}, Heavy Ion Phys. {\bf 15}, 1-80, (2002); arXiv:hep-ph/0001233v3

\bibitem{M-2}
 %Covariant description of kinetic freeze out through a finite time-like layer,
 E. Moln\'ar, L. P. Csernai, V. K. Magas, Zs. I. Lazar, A. Nyiri, and
 K. Tamosiunas, J. Phys. G {\bf 34}, 1901 (2007).
 % arXiv: (nucl-th/0503048)

\bibitem{M-3}
 % Covariant description of kinetic freeze out through a finite spacelike layer,
 E. Moln\'ar, L. P. Csernai, V. K. Magas, A. Nyiri, and K. Tamosiunas,
 Phys. Rev. C {\bf 74}, 024907 (2006).
 %arXiv: (nucl-th/0503047)

\bibitem{Yan12}
Yu-Liang Yan, Yun Cheng, Dai-Mei Zhou, Bao-Guo Dong, Xu Cai,
Ben-Hao Sa, and Laszlo P Csernai, J. Phys. G {\bf 40}, 025102 (2013).

\bibitem{AVC13}
 %Pionic freeze-out hypersurfaces in relativistic nucleus-nucleus collisions
 D. Anchishkin, V. Vovchenko, and L.P. Csernai,
 Phys. Rev. C {\bf 87}, 014906 (2013).

\bibitem{Yun10}
 % Matching stages of heavy-ion collision models
 Y. Cheng, L.P. Csernai, V.K. Magas, B.R. Schlei,  and D. Strottman,
 Phys. Rev. C {\bf 81}, 064910 (2010).

\bibitem{HP12}
 % Particlization in hybrid models
 P. Huovinen, H. Petersen,
 Eur. Phys. J. A {\bf 48}, 171 (2012)

\bibitem{AVY13}
 %Hadronic Reaction Zones in Relativistic Nucleus-Nucleus Collisions
 D. Anchishkin, V. Vovchenko, and S. Yezhov,
 Int. J. Modern Phys. E {\bf 22} 1350042 (2013).

\bibitem{magas}
V.K. Magas, L.P. Csernai, D.D. Strottman, 
Phys. Rev. C {\bf 64}, 014901 (2001), and
Nucl. Phys. A {\bf 712}, 167-204 (2002).

\bibitem{MLisa}
 M.A. Lisa, N.N. Ajitanand, J.M. Alexander, et al.,
 Phys. Lett. B {\bf 496}, 1 (2000);
 M.A. Lisa, U. Heinz, U.A. Wiedemann,
 Phys. Lett. B {\bf 489}, 287 (2000);
 E. Mount, G. Graef, M. Mitrovski, M. Bleicher, M.A. Lisa, 
 Phys. Rev. C {\bf 84},  014908 (2011);
 G. Graef, M. Bleicher, and M. Lisa,
 Phys. Rev. C {\bf 89}, 014903 (2014).

\bibitem{Eyyubova}
 L.P. Csernai, G. Eyyubova, V.K. Magas,
 Phys. Rev. C {\bf 86}, 024912 (2012).




\end{thebibliography}
\end{document}